\documentclass{article}

\usepackage{arxiv}

\usepackage{amsmath}
\usepackage{graphicx}
\usepackage{hyperref}
\usepackage{booktabs}
\usepackage{subcaption}
\usepackage{tikz}
\usepackage{enumitem}
\usepackage{geometry}
\usepackage{svg}
\usepackage{cite}
\usetikzlibrary{calc} 
\usepackage[utf8]{inputenc} 
\usepackage[T1]{fontenc}    
\usepackage{hyperref}       
\usepackage{url}            
\usepackage{booktabs}       
\usepackage{amsfonts}       
\usepackage{nicefrac}       
\usepackage{microtype}      
\usepackage{lipsum}
\usepackage{graphicx}
\graphicspath{ {./images/} }

\title{KACDP: A Highly Interpretable Credit Default Prediction Model}

\author{
 Kun Liu \\
  School of Information\\
  Xi'an University of Finance and Economics\\
  \texttt{17746305707@126.com} \\
   \And
 Jin Zhao \\
  School of Information\\
  Xi'an University of Finance and Economics\\
  \And
}

\begin{document}
\maketitle
\begin{abstract}
In the field of finance, the prediction of individual credit default is of vital importance. However, existing methods face problems such as insufficient interpretability and transparency as well as limited performance when dealing with high-dimensional and nonlinear data.

To address these issues, this paper introduces a method based on Kolmogorov - Arnold Networks (KANs). KANs is a new type of neural network architecture with learnable activation functions and no linear weights, which has potential advantages in handling complex multi-dimensional data.

Specifically, this paper applies KANs to the field of individual credit risk prediction for the first time and constructs the Kolmogorov - Arnold Credit Dedault Predict (KACDP) model. Experiments show that the KACDP model outperforms mainstream credit default prediction models in performance metrics (ROC\_AUC and F1 values). Meanwhile, through methods such as feature attribution scores and visualization of the model structure, the model's decision-making process and the importance of different features are clearly demonstrated, providing transparent and interpretable decision-making basis for financial institutions and meeting the industry's strict requirements for model interpretability.

In conclusion, the KACDP model constructed in this paper exhibits excellent predictive performance and satisfactory interpretability in individual credit risk prediction, providing an effective way to address the limitations of existing methods and offering a new and practical credit risk prediction tool for financial institutions. 
\end{abstract}


\section{Introduction}
\label{sec:introduction}

In today's financial field, individual credit risk prediction has become a crucial part in the risk management of financial institutions. Accurate default prediction can not only help financial institutions significantly reduce losses but also significantly improve the utilization rate of funds, thereby enhancing their competitiveness in the market \cite{HAN2022}\cite{BASTOS2022}. With the rapid development of financial technology, numerous machine learning and deep learning techniques are gradually being widely applied in credit risk assessment. However, the existing various methods inevitably expose certain limitations when dealing with high-dimensional and nonlinear data, among which the problems of interpretability and transparency are the most prominent \cite{ahmed2022artificial}.

Traditional credit risk prediction methods mainly include two categories: statistical models and machine learning models. The typical representative of statistical models, such as Logistic regression \cite{dumitrescu_machine_2022}, has the advantage of being simple and easy to use. However, when dealing with complex data, due to relatively strict assumptions, it is often difficult to effectively capture nonlinear relationships. Machine learning models, such as Random Forest (RF) \cite{malekipirbazari_risk_2015}, Support Vector Machine (SVM) \cite{kim_support_2010}, and Extreme Gradient Boosting Machine (XGBoost) \cite{li_machine_2020}, although they perform relatively well in handling high-dimensional data, their interpretability is relatively poor and it is difficult to provide a clear and transparent decision-making process. Deep learning models, like Multi-Layer Perceptron (MLP) \cite{Duan_2019} and Recurrent Neural Network (RNN) \cite{babaev_et-rnn_2019}, although they have strong expressive ability, in the practical application in the financial field, their black-box characteristics cause the model to severely lack transparency and interpretability, which undoubtedly becomes a major problem in the strictly regulated financial industry \cite{setzu_glocalx_2021}.

Kolmogorov - Arnold Networks (KANs) is a new type of neural network architecture based on the Kolmogorov - Arnold representation theorem, featuring learnable activation functions and no linear weights. When dealing with complex multi-dimensional data, KANs has excellent performance, high accuracy, and interpretability. The core is to represent the multi-dimensional function as a combination of univariate functions, maintaining strong expressive ability while having good interpretability \cite{liu2024kan}.
KANs have advantages in processing financial data because financial data is highly dimensional and nonlinear, and KANs can adapt flexibly; and the financial industry has high requirements for the transparency and interpretability of the model, and KANs can meet them.

This paper aims to comprehensively evaluate the performance and interpretability of KANs in individual credit risk prediction and provide a new method. The innovation lies in the first application of KANs in this field, constructing the Kolmogorov - Arnold Credit Dedault Predict (KACDP) model, and achieving satisfactory results. The specific contributions are as follows:
\setlist [enumerate]{topsep=1pt,parsep=0pt}
\begin {enumerate}
\item Innovation in model application: It is the first time to apply KANs to the field of individual credit risk prediction, introducing a new neural network architecture to this field and enriching the method library.
\item Performance improvement: Through experimental comparisons, it is proved that the KACDP model outperforms traditional statistical models and machine learning models in key performance indicators such as ROC\_AUC and F1 values. The unique network structure and learning mechanism of KANs enable it to better capture the characteristics of high-dimensional and nonlinear data and improve the prediction accuracy.
\item Enhanced interpretability: In-depth exploration of the interpretability of the KACDP model, through methods such as feature importance analysis and model structure visualization, clearly shows the model decision-making process and feature importance, providing transparent and interpretable decision-making basis for financial institutions and meeting industry requirements.
\end {enumerate}

The unique advantages of the KACDP model are expected to solve the limitations of existing methods when dealing with financial data and provide an effective tool for financial institutions.

\section{Related Work}
\label{sec:related_work}
\subsection{Review of Credit Default Prediction Methods}

Personal credit risk prediction is an important issue in the financial field, which is of great significance for the risk management and credit decision-making of financial institutions \cite{HAN2022}\cite{BASTOS2022}. The existing research methods mainly include two categories: traditional statistical methods and machine learning methods.

Traditional statistical methods include Bayesian classification, Logistic regression, Probit regression, and discriminant analysis, etc. Abdou et al. \cite{abdou_would_2019} studied the application of four statistical modeling techniques (discriminant analysis, Logistic regression, multi-layer feedforward neural network, and probabilistic neural network) in credit scoring in the Indian banking industry and found that complex credit scoring models can significantly reduce the default rate. Chi Guotai et al. \cite{CGT_2016} proposed a small business credit rating model based on Probit regression. By screening and optimizing the rating indicators, the prediction accuracy of the model and the discrimination ability of the default state were improved. Dumitrescu et al. \cite{dumitrescu_machine_2022} proposed an improved Logistic regression model - Penalized Logistic Tree Regression (PLTR), which improved the prediction performance of credit scoring by combining the nonlinear effect of decision trees while retaining the interpretability of Logistic regression. Mahbobi et al. \cite{mahbobi_credit_2023} incorporated the K-Nearest Neighbor method into the proposed combined model, effectively improving the accuracy of credit default prediction. Mancisidor et al. \cite{mancisidor_deep_2020} proposed a Bayesian model combining Gaussian mixture models and auxiliary variables to handle the rejection inference problem in credit scoring, aiming to improve the classification accuracy through semi-supervised learning.

Although traditional statistical methods are widely used in the field of credit default prediction, their prediction performance is low and they cannot fit the data well. Therefore, machine learning methods have attracted people's attention. Yao et al. \cite{yao_support_2015} used the SVR technique to predict the default loss rate of corporate bonds and improved the prediction accuracy through improved algorithms. Pang et al. \cite{pang_borrowers_2021} proposed a borrower credit quality scoring model based on Extreme Learning Machine (ELM), Fuzzy C-Means algorithm (FCM), and confusion matrix to evaluate the default probability and default loss rate. Yang et al. \cite{yang_deep_2023} proposed an improved Deep Neural Network algorithm (HDNN), which effectively solved the overfitting problem in high-dimensional enterprise credit risk prediction by adding L1 regularization and L2 constraints to the batch normalization layer. Lee et al. \cite{lee_graph_2021} proposed a credit default prediction model based on Graph Convolutional Network (GCN), which significantly improved the prediction performance by integrating three types of virtual distance information between borrowers. In addition, Reinforcement Learning (DL), Recurrent Neural Network (RNN) \cite{babaev_et-rnn_2019}, Long Short-Term Memory Network (LSTM) \cite{shen_new_2021} have also been used for credit default prediction with temporal features.

However, a single machine learning model is prone to fall into local optimum during training and has poor generalization ability. Ensemble learning makes decisions by integrating the synthesis of basic models and has good stability and generalization ability. Malekipirbazari and Aksakalli \cite{malekipirbazari_risk_2015} used the Random Forest (RF) method to predict the borrower status on the social lending platform Lending Club, and the effect was better than the traditional FICO credit score and LC grade. Xu et al. \cite{xu_efficient_2023} proposed an efficient fraud detection method based on Deep Boosted Decision Tree (DBDT), which improved the classification performance on imbalanced datasets by combining the AUC maximization strategy. In addition, Lightweight Gradient Boosting Machine (LightGBM), Extreme Gradient Boosting Machine (XGBoost), and Adaptive Boosting (AdaBoost) \cite{li_machine_2020} have also been used for credit default prediction.

In May 2018, the European Parliament approved the EU Data Protection Law (EU Data Protection Law). The EU Data Protection Law has strict restrictions on automated black box models.
Since the European Parliament passed the EU Data Protection Law, uninterpretable ML models have been restricted by financial institutions for credit default prediction \cite{setzu_glocalx_2021}. Therefore, scholars have shifted their attention to interpretable models or "black box" models with interpretable methods. Currently, the most popular machine learning interpretability methods are SHAP (SHapley Additive exPlanations) and LIME (Local Interpretable Model-agnostic Explanations). Park et al. \cite{Park_2021} explored the interpretability issue of machine learning models in bankruptcy prediction and used the LIME method to improve the transparency and interpretability of the model to better understand and trust the prediction results. Mahajan et al. \cite{Mahajan_2023} used the SHAP technique to explain the bankruptcy prediction models of companies in Taiwan and Poland and analyzed the contributions of different features in false positive cases, thereby providing an in-depth understanding of the model's prediction errors.

\subsection{Related Work on Kolmogorov-Arnold Networks (KANs)}
Recently, inspired by the Kolmogorov-Arnold representation theorem, Liu et al. \cite{liu2024kan} proposed a new neural network architecture KANs (Kolmogorov-Arnold Network), which demonstrated higher accuracy and interpretability in small-scale AI+Science tasks by using learnable activation functions on the edges instead of the fixed activation functions in traditional MLPs. Subsequent studies have applied KANs to various fields. Bresson et al. \cite{bresson2024kagnns} combined Kolmogorov-Arnold networks and Graph Neural Networks (GNNs) to propose a new graph learning model KAGNNs, which is applied to graph data. Wang et al. \cite{wang2024kolmogorovarnoldinformedneural} proposed a physics-informed deep learning framework KINN based on Kolmogorov-Arnold networks, which can combine physical laws in forward and inverse problem solving to improve the accuracy and generalization ability of the model. Liu et al. \cite{liu2024ikanglobalincrementallearning} proposed an incremental learning method iKAN based on Kolmogorov-Arnold networks (KAN), specifically for Human Activity Recognition (HAR). Yang et al. \cite{yang2024endowinginterpretabilityneuralcognitive} proposed a neural cognitive diagnosis model KAN2CD based on Kolmogorov-Arnold networks (KAN), which enhanced the interpretability of the model by replacing the traditional Multi-Layer Perceptron (MLP), enabling the model to more clearly show students' mastery of knowledge concepts. Handal et al. \cite{handal2024kanopdataefficientoptionpricing} proposed a data-efficient option pricing model KANOP based on Kolmogorov-Arnold networks (KAN), which combined the traditional Least Squares Monte Carlo (LSMC) algorithm when dealing with American-style option pricing, improving the accuracy and data efficiency of the model.

In addition, KAN is also used in combination with other technologies to improve the accuracy in specific fields or be applicable to specific data. Chen et al. \cite{chen2024sckansformerfinegrainedclassificationbone} proposed a fine-grained bone marrow cell classification model named SCKansformer, which combines the Kansformer encoder, SCConv encoder, and global-local attention mechanism, effectively solving the problems of poor feature expression ability, poor interpretability, and redundant feature extraction in the processing of high-dimensional microscopic image data, and improving the accuracy and efficiency of classification. Bodner et al. \cite{bodner2024convolutionalkolmogorovarnoldnetworks} proposed a novel Convolutional Kolmogorov-Arnold Network (Convolutional KANs), which introduced the nonlinear activation functions in the Kolmogorov-Arnold network into the convolution operation and constructed a new network layer. Experimental results show that Convolutional KANs outperforms traditional Convolutional Neural Networks (CNNs) on multiple benchmark datasets.

\section{Methodology}
\label{sec:methodology}
In this section, we first introduce the core principle of the KANs introduced in this paper in \ref{subsec:kan}, then construct the KACDP model based on the core principle of KANs in \ref{subsec:KACDP}, and finally introduce the benchmark machine learning models for credit default prediction in \ref{subsec:base_ml}.
\subsection{Kolmogorov-Arnold Networks (KANs)}
\label{subsec:kan}

Kolmogorov-Arnold Networks (KANs) is a new type of neural network architecture based on the Kolmogorov-Arnold representation theorem. The core idea of KANs is to represent multi-dimensional functions as combinations of univariate functions, which enables the model to have good interpretability while maintaining high expressive ability.

1. \textbf{Kolmogorov-Arnold Representation Theorem}: The Kolmogorov-Arnold representation theorem states that any continuous function can be represented as a combination of several univariate functions. This theory provides a theoretical basis for the design of KANs.
\[
f(x) = \sum_{q=1}^{2n+1} \Phi_q\left(\sum_{p=1}^n \phi_{q,p}(x_p)\right)
\]
where \(\phi_{q,p}\) and \(\phi_q\) are learnable univariate functions that act on input features and intermediate features, respectively.

2. \textbf{Deep KANs}: Based on the Kolmogorov-Arnold representation theorem, two-layer KANs can be represented. If the KAN layer is defined as \(\Phi = \{\phi_{q,p}\}\), where \(p = 1, 2, \cdots, n_{in}\) and \(q = 1, 2, \cdots, n_{out}\) represent the input dimension and output dimension, respectively.
Then the calculation process of the KANs with layer number \(L\) can be expressed as:
\[
\text{KAN}(x) = (\Phi_{L-1} \circ \Phi_{L-2} \circ...\circ\Phi_0)(x),
\]
where each \(\Phi_i\) is a KAN layer.

3. \textbf{Learnable Activation Functions}: For each learnable univariate activation function \(\phi(x)\), it contains a base function \(b(x)\) (similar to a residual connection) and a B-spline function \(\text{spline}(x)\):
\[
\phi(x) = w_b b(x) + w_s \text{spline}(x)
\]
In most cases, we set
\[
b(x) = \text{silu}(x) = \frac{x}{1 + e^{-x}},
\]
where \(\text{spline}(x)\) is parameterized as a linear combination of B-splines:
\[
\text{spline}(x) = \sum_i c_i B_i(x),
\]
where \(c_i\), \(w_b\), and \(w_s\) are trainable parameters.

\subsection{Construction of the KACDP Model}
\label{subsec:KACDP}
The KACDP model includes two models with different structures, KACDP\_OP (Kolmogorov - Arnold Credit Dedault Predict Optimal preformance, representing the best performance) and KACDP\_OI (Kolmogorov - Arnold Credit Dedault Predict Optimal interpretability, representing the strongest interpretability). They differ in network structure, construction purpose, and some hyperparameter settings, but are all constructed based on the Kolmogorov - Arnold Networks (KANs) theory.
For the KACDP\_OP model, its network structure is width = [10, 4, 1], including 10 input features, a hidden layer with four neurons, and an output layer. When determining the number of neurons in the hidden layer, after multiple rounds of experiments, considering the balance between model complexity and performance, and comparing the performance of different numbers of neurons on the validation dataset, it was finally determined that four neurons can balance the expressive ability and avoid overfitting to achieve good performance. The 10 features of the input layer cover multiple key aspects of personal credit risk, including the borrower's basic information (such as age), credit history indicators (such as the number of times in different past overdue periods, credit card usage rate, etc.) and economic status indicators (such as the ratio of total debt to monthly income, monthly income, etc.). These feature selections are based on an in-depth understanding of the credit risk assessment field and relevant research references \cite{HAN2024}.

\begin{figure}[htbp]
    \centering
    \begin{minipage}{0.48\textwidth}
        \centering
       \includegraphics[width=\textwidth]{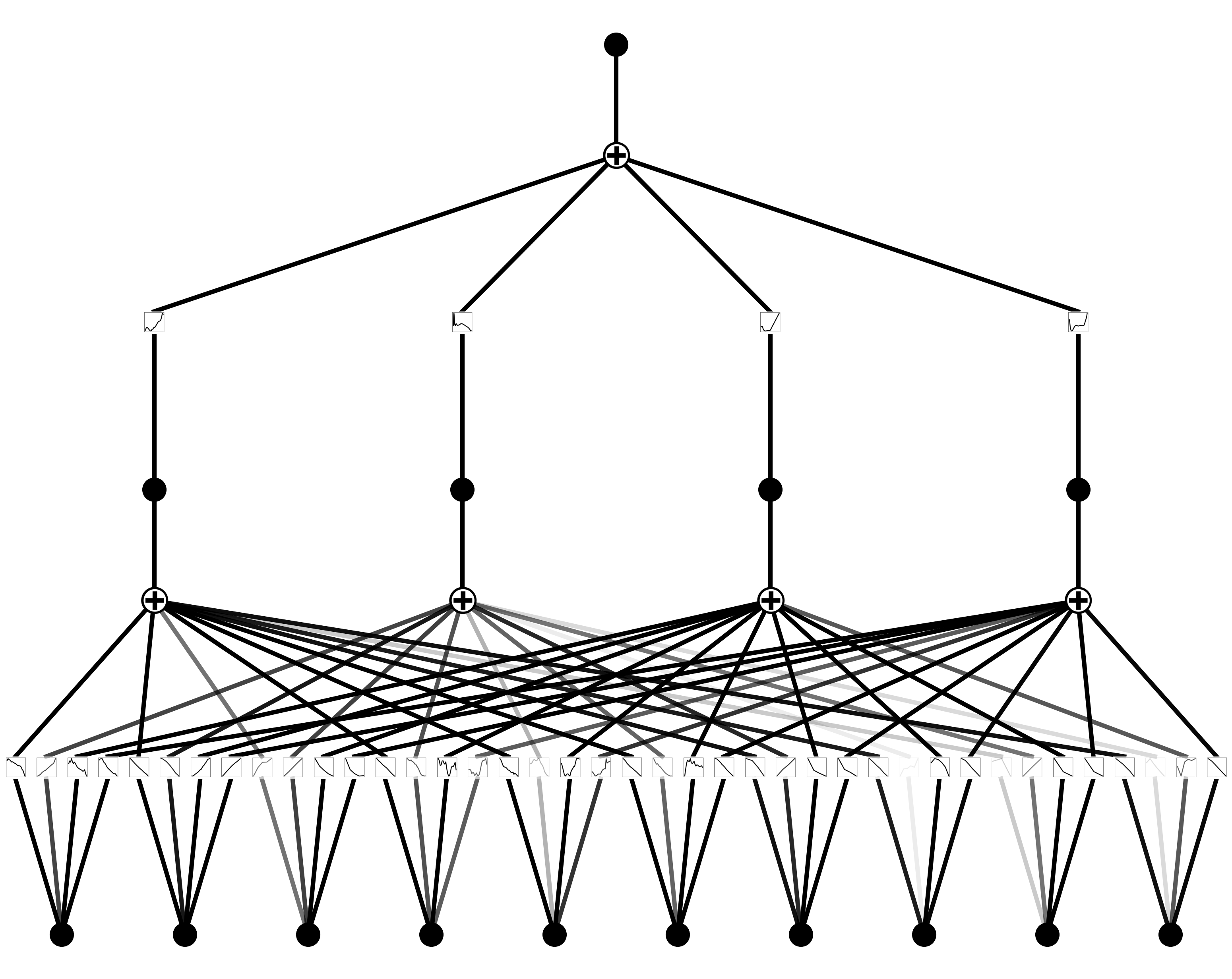}
        \subcaption{Structure diagram of the KACDP\_OP model}
        \label{fig:modelOP_structure}
    \end{minipage}
    \hfill
    \hfill
    \begin{minipage}{0.48\textwidth}
        \centering
        \includegraphics[width=\textwidth]{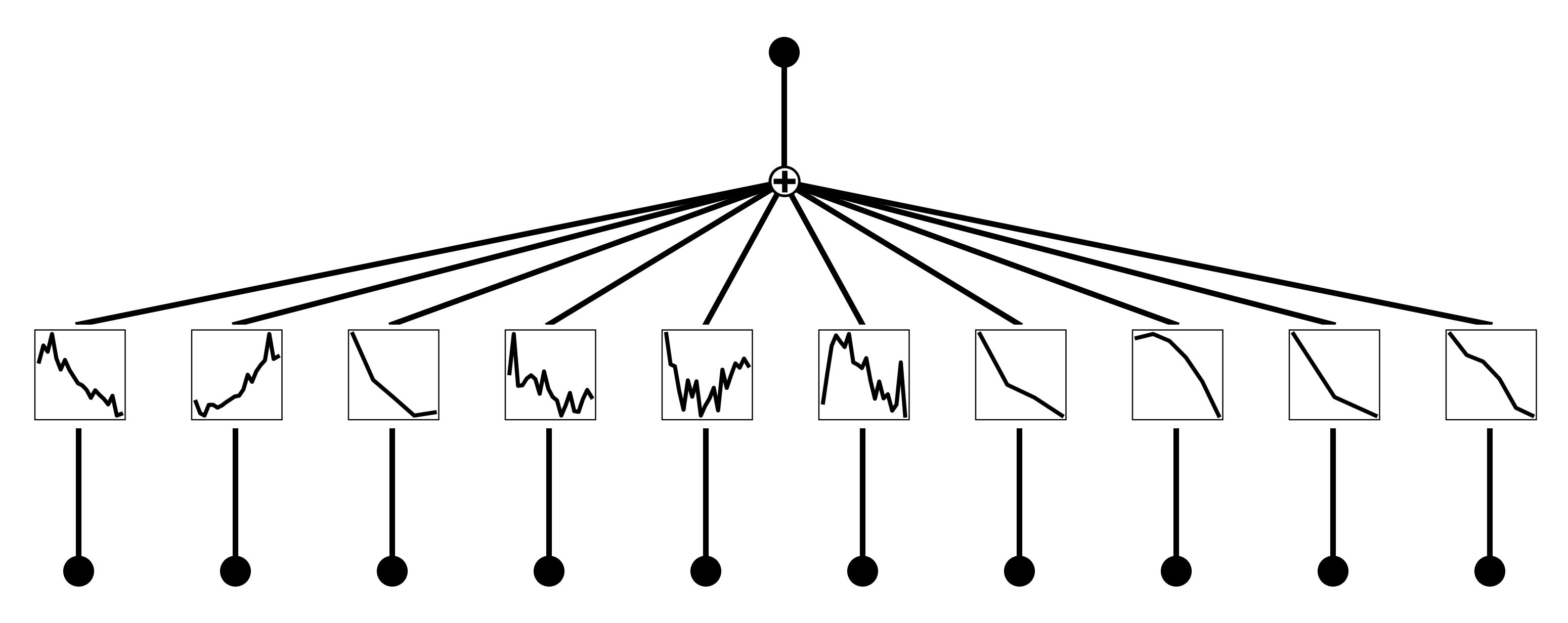}
        \subcaption{Structure diagram of the KACDP\_OI model}
        \label{fig:modelOI_structure}
    \end{minipage}
    \caption{Structure diagrams of the KACDP\_OP and KACDP\_OI models}
    \label{fig:OPOI}
\end{figure}

The model training uses the Adam optimizer, the batch size is set to -1 (full batch training), and the learning rate is 0.1. This learning rate is determined by comprehensively considering the convergence speed and stability of the model to ensure that the training process is stable and can fully utilize the information of the data samples to help the model learn the patterns and rules of the data. The relevant parameters of B-spline are set as: grid (controlling the number of grids) is 30, and \(k\) (approximation order) is 4. These parameters are crucial for the model to accurately fit the complex patterns of the data. Reasonable settings make the KACDP\_OP model perform well in the performance evaluation. At the same time, the KACDP\_OP model can be visualized (as shown in the subgraph \ref{fig:modelOP_structure} in Figure \ref{fig:OPOI}), and the network structure and the connection relationship of each layer can be visually displayed through the model structure diagram, which is helpful for understanding the working principle of the model.

The KACDP\_OI model focuses on interpretability, and the network structure is set to width = [10, 1], including only the input layer and the output layer. This simplified structure helps to understand the decision-making process intuitively. The input layer also contains 10 important features related to personal credit risk, and the selection basis is the same as that of the KACDP\_OP model. Model training also uses the Adam optimizer, the batch size is set to -1, and the learning rate is 0.1. Despite the simple structure, by setting the hyperparameters reasonably, this model achieves satisfactory performance while maintaining good interpretability. In terms of B-spline parameters, grid is set to 80 and \(k\) is set to 4. Compared with the KACDP\_OP model, the different grid values reflect the differences in data fitting requirements under different structures. By adjusting the parameters, the KACDP\_OI model can balance the prediction performance and interpretability on the basis of the simplified structure. The KACDP\_OI model can also be visualized (as shown in the subgraph \ref{fig:modelOI_structure} in Figure \ref{fig:OPOI}), and the input layer features can be clearly displayed through its model structure diagram to show how the output results are obtained through simple processing, reflecting the characteristics of interpretability.
Through the above construction process, the KACDP\_OP and KACDP\_OI models have outstanding characteristics in terms of performance and interpretability, respectively, providing a basis for subsequent experimental analysis and comparison.
\subsection{Benchmark Machine Learning Methods}
\label{subsec:base_ml}
\subsubsection{Support Vector Machine (SVM)}
Support Vector Machine (SVM) is a powerful machine learning algorithm widely used in classification and regression problems. Its core idea is to find an optimal hyperplane in the feature space to maximize the margin between samples of different classes. SVM finds this hyperplane by solving the following optimization problem:
\[
\min_{\omega, \beta, \xi} \frac{1}{2} \|\omega\|^2 + C \sum_{i=1}^N \xi_i
\]
\[
\text{subject to } y_i (\omega^T \phi(x_i) + \beta) \geq 1 - \xi_i, \quad \xi_i \geq 0, \quad i = 1, 2, \ldots, N
\]
where \(\omega\) is the weight vector, \(\beta\) is the bias term, \(\phi(x_i)\) is the kernel function that maps the input features to a high-dimensional space, \(\xi_i\) is the slack variable used to handle inseparable cases, and \(C\) is the regularization parameter that controls the complexity of the model. SVM improves the generalization ability of the model by maximizing the margin, making it perform well when dealing with small sample datasets.

\subsubsection{Random Forest (RF)}
Random Forest (RF) is a classification and regression method based on ensemble learning, which makes predictions by constructing multiple decision trees and aggregating their results. The core idea of RF is to use the bagging method to extract multiple subsamples from the original dataset, and then train a decision tree on each subsample. The final prediction result is determined by majority voting (for classification tasks) or averaging (for regression tasks). The advantages of RF lie in its ability to reduce the variance of individual decision trees, improve the stability and prediction performance of the model. In addition, RF can handle high-dimensional data and missing values and has good interpretability.

\subsubsection{XGBoost}
XGBoost is an efficient machine learning algorithm based on Gradient Boosting Decision Tree (GBDT), which is widely used in recommendation systems, demand forecasting, and data mining. The basic idea of XGBoost is to gradually reduce the loss function of the model by iteratively adding new decision trees. Compared with the traditional GBDT, XGBoost introduces several improvements, including using the second-order Taylor expansion to approximate the loss function to obtain more accurate gradient information; introducing regularization terms to prevent overfitting; using block storage and parallel computing to improve training efficiency. These improvements make XGBoost perform well when dealing with large-scale datasets, with high prediction accuracy and computational efficiency.

\section{Experimental Results and Analysis}
\label{sec:results}
In this section, we first introduce the data sources and performance evaluation metrics of the models used in this study in \ref{subsec:dataandmetric}, then show the performance comparison of KACDP\_OP, KACDP\_OI, Logistic Regression, XGBoost, and Support Vector Machine in terms of ROC\_AUC and F1 values in \ref{subsec:performance_comparison}, and finally demonstrate the interpretability of the KACDP model through feature attribution scores and model structure visualization in \ref{subsec:interpretability}. Additionally, in the Appendix \ref{subsec:hyperparameters}, the hyperparameter sensitivity analysis of the KACDP model is introduced as a supplementary experiment for researchers' reference.
\subsection{Data Source and Model Evaluation Metrics}
\label{subsec:dataandmetric}
\subsubsection{Data Source}
\label{subsec:data_description}
This study conducts experiments using the publicly available personal credit dataset GMSC (Give Me Some Credit). This dataset contains 223,958 records, each representing the credit information of a borrower. The GMSC dataset is derived from the personal credit data of a large bank and includes the basic information, credit history, and loan records of borrowers. There are 10 features and one target variable:
\setlist[itemize]{itemsep=0pt,topsep=0pt,partopsep=0pt,parsep=0pt}
\begin{itemize}
    \item \textbf{SeriousDlqin2yrs}: Whether the borrower has had a serious default within the past two years (target variable, 1 indicates default, 0 indicates no default).
    \item \textbf{RevolvingUtilizationOfUnsecuredLines}: The utilization rate of the borrower's credit card, that is, the ratio of the outstanding credit card balance to the credit limit.
    \item \textbf{age}: The age of the borrower.
    \item \textbf{NumberOfTime30-59DaysPastDueNotWorse}: The number of times the borrower was overdue for 30-59 days in the past record.
    \item \textbf{DebtRatio}: The ratio of the borrower's total debt to monthly income.
    \item \textbf{MonthlyIncome}: The monthly income of the borrower.
    \item \textbf{NumberOfOpenCreditLinesAndLoans}: The number of open credit accounts and loans of the borrower.
    \item \textbf{NumberOfTimes90DaysLate}: The number of times the borrower was overdue for 90 days in the past record.
    \item \textbf{NumberRealEstateLoansOrLines}: The number of real estate loans or credit lines of the borrower.
    \item \textbf{NumberOfTime60-89DaysPastDueNotWorse}: The number of times the borrower was overdue for 60-89 days in the past record.
    \item \textbf{NumberOfDependents}: The number of family members the borrower needs to support.
\end{itemize}
This dataset is widely used in the research of personal credit default prediction and has high representativeness and reliability \cite{HAN2024}\cite{gong2022hybrid}\cite{sharma2023credit}. In addition, the data preprocessing and feature engineering in this article are based on \cite{HAN2024}.
\subsubsection{Evaluation Metrics}
\label{subsec:evaluation_metrics}
In the real world, the situation of personal credit default is usually extremely unbalanced. When predicting defaults, the two indicators of precision and recall must be balanced. Therefore, to objectively evaluate the predictive performance of the model, this article uses two indicators: the area under the ROC curve (ROC\_AUC) and the F1 score (F1 Score).

ROC\_AUC (ROC Area Under the Curve) refers to the area under the Receiver Operating Characteristic Curve (ROC Curve). The ROC curve is generated by plotting the relationship between the True Positive Rate (TPR) and the False Positive Rate (FPR). The AUC indicator considers the model's ability to distinguish positive and negative examples and can provide a reasonable evaluation even in the case of unbalanced samples. The value of AUC ranges from 0 to 1, and the closer the value is to 1, the better the classification performance of the model.

The F1 score is the harmonic mean of precision and recall, effectively balancing these two indicators. The calculation formula of the F1 score is:
\[
\text{F1} = 2 \times \frac{\text{Precision} \times \text{Recall}}{\text{Precision} + \text{Recall}}
\]
Where:
Precision: Represents the proportion of samples predicted as positive that are actually positive.
\[
\text{Precision} = \frac{\text{TP}}{\text{TP} + \text{FP}}
\]
Recall: That is, TPR, represents the proportion of samples that are actually positive and are correctly predicted as positive.
\[
\text{Recall} = \frac{\text{TP}}{\text{TP} + \text{FN}}
\]

The value of the F1 score ranges from 0 to 1, and the closer the value is to 1, the better the predictive performance of the model.

\subsection{Performance Comparison}
\label{subsec:performance_comparison}

To evaluate the performance of KACDP in the task of personal credit risk prediction. We selected several representative traditional models that have been widely used in academia and industry as comparative references, including Logistic Regression, XGBoost, and Support Vector Machine \cite{li_machine_2020}\cite{dumitrescu_machine_2022}\cite{yao_support_2015}. These traditional models have undergone extensive experimental verification in many previous studies, and their performance and application scope are well known in the academic community. To ensure the scientific and accuracy of the comparison, we directly cited the research results of Han et al. \cite{HAN2024} because they conducted relevant experiments on the same dataset as our study, which provides us with a relatively fair and reliable basis for comparison.

It can be clearly seen from the data in Table \ref{tab:performance_comparison} that KACDP\_OP shows significant advantages in the ROC\_AUC indicator, with a value reaching 0.8670, which is significantly higher than the traditional Logistic Regression (0.8503), XGBoost (0.8634), and Support Vector Machine (0.8555). At the same time, in the F1 value indicator, both KACDP\_OP and KACDP\_OI achieved a score of 0.9675, also superior to other traditional models. For more performance analysis of KACDP, see \ref{subsec:hyperparameters}.

\begin{table}[htbp]
    \centering
    \caption{Performance comparison of different models in terms of ROC\_AUC and F1 indicators}
    \label{tab:performance_comparison}
    \begin{tabular}{lcc}
        \toprule
        Model & ROC\_AUC & F1 Value \\
        \midrule
        Logistic Regression & 0.8503 & 0.9665 \\
        XGBoost & 0.8634 & 0.9669 \\
        Support Vector Machine & 0.8555 & 0.9665 \\
        \textbf{KACDP\_OP} & \textbf{0.8670} & \textbf{0.9675} \\
        KACDP\_OI & 0.8640 & 0.9675 \\
        \bottomrule
    \end{tabular}
    
\end{table}

\subsection{Interpretability Analysis}
\label{subsec:interpretability}

Interpretability is an important feature of financial models, especially in credit risk prediction. Financial institutions need to be able to understand the decision-making process of the model in order to make transparent and responsible decisions under legal and regulatory requirements. KANs, as a new type of neural network architecture, has good interpretability while maintaining high expressive ability. This section will detail the interpretability characteristics of KANs in personal credit risk prediction and demonstrate its interpretability through feature importance analysis and visualization.

\subsubsection{Feature Importance Analysis}
\label{subsec:feature_importance}

Feature importance analysis is one of the common methods for evaluating the interpretability of the model. By calculating the influence degree of each feature on the model's prediction results, we can understand which features play a key role in the model's decision-making. KANs quantifies the importance of input features by iteratively calculating feature attribution scores \cite{liu2024kan2}.

\begin{figure}[htbp]
    \centering
    \includegraphics[width=0.8\textwidth]{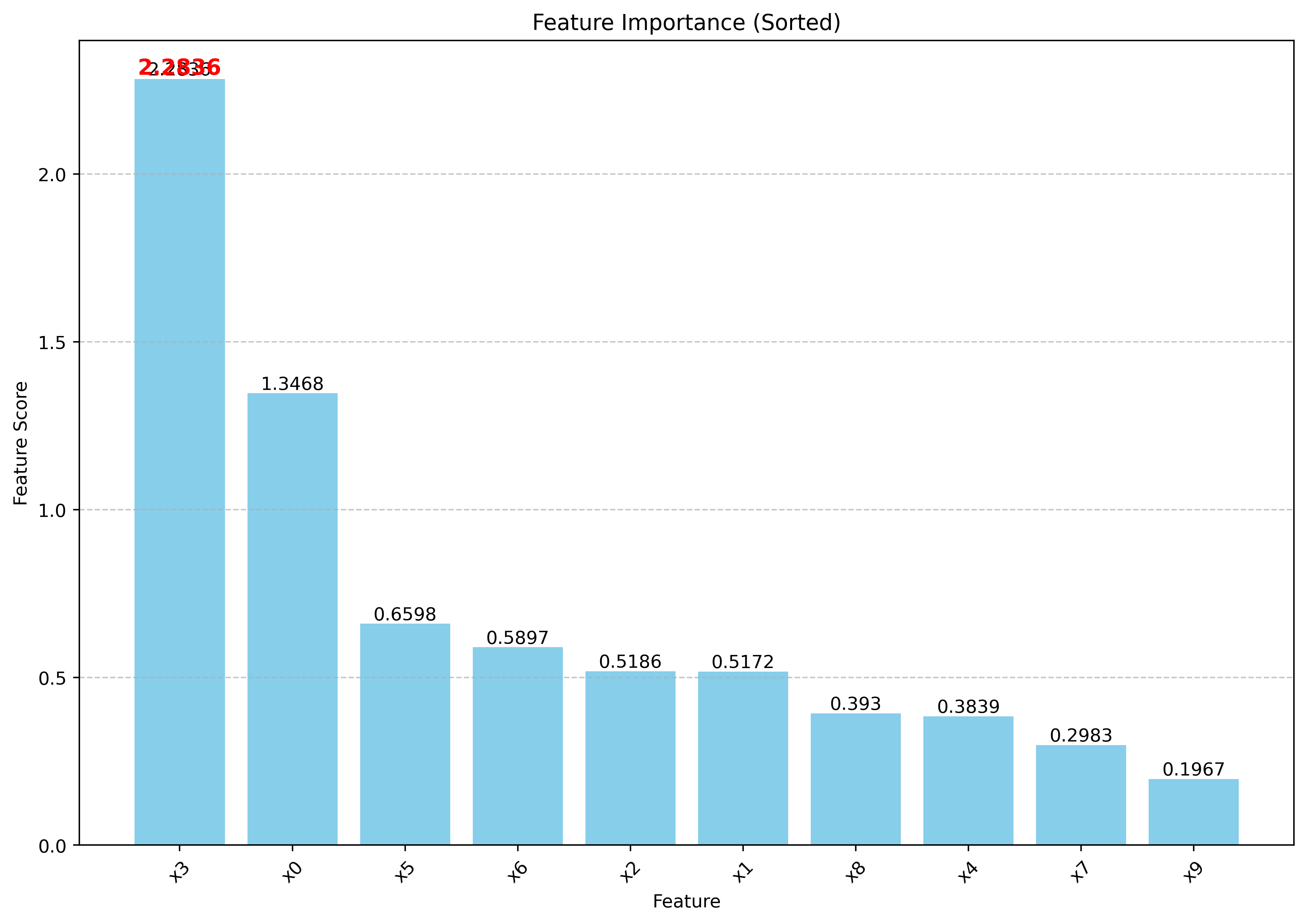}
    \caption{Feature importance analysis of the KACDP\_OP model}
    \label{fig:Feature_importance}
\end{figure}

Figure~\ref{fig:Feature_importance} shows the importance scores of each feature in the KACDP\_OP model. It can be seen from the figure that KACDP can clearly identify the features that have the greatest impact on credit risk prediction. For example: X3 (the ratio of the borrower's total debt to monthly income) shows higher importance, with a score of 2.2836. Followed by x0 (the utilization rate of the borrower's credit card), with a score of 1.3468. These features may play a key role in the model's decision-making process and have a greater impact on the prediction results of credit risk. While features like x9 (the number of family members the borrower needs to support) have a relatively low score of 0.1967, and may be relatively less important in the model.

The features and their meanings in Figure~\ref{fig:feature_importance} are as follows:
- \textbf{x0}: The utilization rate of the borrower's credit card
- \textbf{x1}: The age of the borrower
- \textbf{x2}: The number of times the borrower was overdue for 30-59 days in the past record
- \textbf{x3}: The ratio of the borrower's total debt to monthly income
- \textbf{x4}: The monthly income of the borrower
- \textbf{x5}: The number of open credit accounts and loans of the borrower
- \textbf{x6}: The number of times the borrower was overdue for 90 days in the past record
- \textbf{x7}: The number of real estate loans or credit lines of the borrower
- \textbf{x8}: The number of times the borrower was overdue for 60-89 days in the past record, and there is no worse overdue record
- \textbf{x9}: The number of family members the borrower needs to support

\subsubsection{Visualization of Model Structure and Decision Path}
\label{subsec:model_visualization}
In the field of machine learning, the visualization of the model structure is of crucial significance for understanding the working principle and decision-making process of the model. For the KACDP model, due to its unique characteristic of being based on the superposition of simple univariate functions, we can clearly show its decision-making process by drawing the model structure diagram.

\begin{figure}[htbp]
    \centering
    \begin{minipage}{\textwidth}
        \centering
       \includegraphics[width=\textwidth]{modelOI.png}
        \subcaption{Structure diagram of the KACDP\_OI model}
        \label{fig:model_structure}
    \end{minipage}
    \hfill
    \hfill
    \begin{minipage}{\textwidth}
        \centering
        \includegraphics[width=\textwidth]{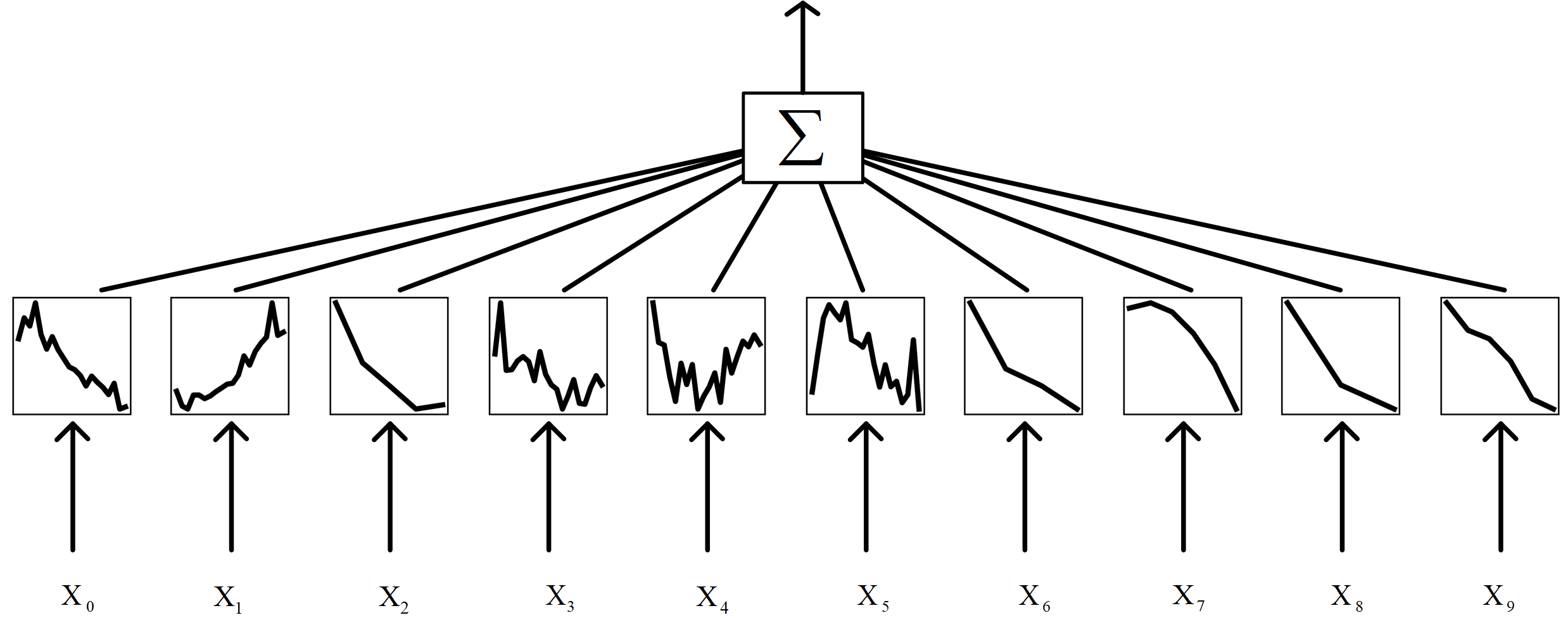}
        \subcaption{Decision path diagram of the KACDP\_OI model}
        \label{fig:decision path}
    \end{minipage}
    \caption{Structure diagram and decision path diagram of the KACDP\_OI model}
    \label{fig:ab}
\end{figure}

Here, we select KACDP\_OI to draw the model structure diagram, and the result is shown in Figure \ref{fig:ab}. In the subfigure \ref{fig:model_structure} of this figure, the structure diagram of KACDP\_OI is detailed. This structure diagram mainly includes two key parts: the input layer and the output layer. Starting from the input layer, each input variable has to be processed by the trained activation function. These activation functions are learned by the model during the training process, and their processing methods for each variable are unique and targeted. After being processed by the activation function, the variables will perform a simple summation operation to finally obtain the output result.

The advantage of this structural form is that it can very clearly depict the decision path of each sample in the model. For each input sample, we can track the entire process from the input layer through the activation function processing to the summation output based on this model structure. This provides an intuitive basis for us to deeply understand how the model makes decisions for different samples. A clearer decision path is shown in the subfigure \ref{fig:decision path}. The detailed explanations of x0-x9 are in \ref{fig:feature_importance}.

\section{Conclusion}
This paper focuses on the application of Kolmogorov - Arnold Networks (KANs) in personal credit risk prediction. The main goal is to evaluate its performance and interpretability and compare it with traditional methods. By constructing the Kolmogorov - Arnold Credit Dedault Predict (KACDP) model, we have reached the following important conclusions.

In terms of performance evaluation, the KACDP model has demonstrated excellent performance in the task of personal credit risk prediction. Compared with traditional statistical models (such as Logistic Regression) and machine learning models (such as Random Forest, Support Vector Machine, Extreme Gradient Boosting Machine) and deep learning models (such as Multi-Layer Perceptron, Recurrent Neural Network), the KACDP model has performed outstandingly in two key performance indicators, ROC\_AUC and F1 values. This is attributed to the unique network structure and learning mechanism of KANs, its learnable activation functions and no linear weights, as well as the core design concept of representing multi-dimensional functions as combinations of univariate functions, enabling the model to better capture the complex patterns in high-dimensional and nonlinear data, thereby improving the accuracy of prediction.

In terms of interpretability, the KACDP model also performs prominently. Through methods such as feature importance analysis, we have clearly demonstrated the model's utilization of each feature and the importance of different features in predicting credit risk. This forms a sharp contrast to the black-box characteristics of traditional deep learning models. The KACDP model can provide more transparent and interpretable decision-making basis for financial institutions, meeting the strict requirements of the financial industry for model transparency and interpretability.

In conclusion, this paper is the first to apply KANs to the field of personal credit risk prediction and construct the KACDP model, which has achieved satisfactory results in both performance and interpretability. It provides a new and effective credit risk prediction tool for financial institutions and is expected to solve the limitations of existing methods when dealing with financial data.

\bibliographystyle{unsrt}  
\bibliography{references}  

\begin{thebibliography}{10}

\bibitem{HAN2022}
Liang Han, Guijun Yang, Xiaodong Yang, Xiaoyu Song, Bo~Xu, Zhenhai Li, Jintao Wu, Hao Yang, and Jianwei Wu.
\newblock An explainable xgboost model improved by smote-enn technique for maize lodging detection based on multi-source unmanned aerial vehicle images.
\newblock {\em Computers and Electronics in Agriculture}, 194:106804, 2022.

\bibitem{BASTOS2022}
João~A. Bastos and Sara~M. Matos.
\newblock Explainable models of credit losses.
\newblock {\em European Journal of Operational Research}, 301(1):386--394, 2022.

\bibitem{ahmed2022artificial}
Imran Ahmed, Gwanggil Jeon, and Francesco Piccialli.
\newblock From artificial intelligence to explainable artificial intelligence in industry 4.0: a survey on what, how, and where.
\newblock {\em IEEE Transactions on Industrial Informatics}, 18(8):5031--5042, 2022.

\bibitem{dumitrescu_machine_2022}
Elena Dumitrescu, Sullivan Hué, Christophe Hurlin, and Sessi Tokpavi.
\newblock Machine learning for credit scoring: {Improving} logistic regression with non-linear decision-tree effects.
\newblock {\em European Journal of Operational Research}, 297(3):1178--1192, March 2022.

\bibitem{malekipirbazari_risk_2015}
Milad Malekipirbazari and Vural Aksakalli.
\newblock Risk assessment in social lending via random forests.
\newblock {\em Expert Systems with Applications}, 42(10):4621--4631, June 2015.

\bibitem{kim_support_2010}
Hong~Sik Kim and So~Young Sohn.
\newblock Support vector machines for default prediction of {SMEs} based on technology credit.
\newblock {\em European Journal of Operational Research}, 201(3):838--846, March 2010.

\bibitem{li_machine_2020}
Jing-Ping Li, Nawazish Mirza, Birjees Rahat, and Deping Xiong.
\newblock Machine learning and credit ratings prediction in the age of fourth industrial revolution.
\newblock {\em Technological Forecasting and Social Change}, 161:120309, December 2020.

\bibitem{Duan_2019}
Jing Duan.
\newblock Financial system modeling using deep neural networks (dnns) for effective risk assessment and prediction.
\newblock {\em Journal of the Franklin Institute}, 356(8):4716--4731, 2019.

\bibitem{babaev_et-rnn_2019}
Dmitrii Babaev, Maxim Savchenko, Alexander Tuzhilin, and Dmitrii Umerenkov.
\newblock E.{T}.-{RNN}: {Applying} {Deep} {Learning} to {Credit} {Loan} {Applications}.
\newblock In {\em Proceedings of the 25th {ACM} {SIGKDD} {International} {Conference} on {Knowledge} {Discovery} \& {Data} {Mining}}, pages 2183--2190, Anchorage AK USA, July 2019. ACM.

\bibitem{setzu_glocalx_2021}
Mattia Setzu, Riccardo Guidotti, Anna Monreale, Franco Turini, Dino Pedreschi, and Fosca Giannotti.
\newblock {GLocalX} - {From} {Local} to {Global} {Explanations} of {Black} {Box} {AI} {Models}.
\newblock {\em Artificial Intelligence}, 294:103457, May 2021.

\bibitem{liu2024kan}
Ziming Liu, Yixuan Wang, Sachin Vaidya, Fabian Ruehle, James Halverson, Marin Soljačić, Thomas~Y. Hou, and Max Tegmark.
\newblock Kan: Kolmogorov-arnold networks, 2024.

\bibitem{abdou_would_2019}
Hussein~A. Abdou, Shatarupa Mitra, John Fry, and Ahmed~A. Elamer.
\newblock Would two-stage scoring models alleviate bank exposure to bad debt?
\newblock {\em Expert Systems with Applications}, 128:1--13, August 2019.

\bibitem{CGT_2016}
Guotai Chi, Yajing Zhong, and Baofeng Shi.
\newblock The debt rating for small enterprises based on probit regression.
\newblock {\em Journal of Management Sciences in China}, 19(06):136--156, 2016.

\bibitem{mahbobi_credit_2023}
Mohammad Mahbobi, Salman Kimiagari, and Marriappan Vasudevan.
\newblock Credit risk classification: an integrated predictive accuracy algorithm using artificial and deep neural networks.
\newblock {\em Annals of Operations Research}, 330(1-2):609--637, November 2023.

\bibitem{mancisidor_deep_2020}
Rogelio~A. Mancisidor, Michael Kampffmeyer, Kjersti Aas, and Robert Jenssen.
\newblock Deep generative models for reject inference in credit scoring.
\newblock {\em Knowledge-Based Systems}, 196:105758, May 2020.

\bibitem{yao_support_2015}
Xiao Yao, Jonathan Crook, and Galina Andreeva.
\newblock Support vector regression for loss given default modelling.
\newblock {\em European Journal of Operational Research}, 240(2):528--538, January 2015.

\bibitem{pang_borrowers_2021}
Professor~Sulin Pang, Xianyan Hou, and Lianhu Xia.
\newblock Borrowers’ credit quality scoring model and applications, with default discriminant analysis based on the extreme learning machine.
\newblock {\em Technological Forecasting and Social Change}, 165:120462, April 2021.

\bibitem{yang_deep_2023}
Mei Yang, Ming~K. Lim, Yingchi Qu, Xingzhi Li, and Du~Ni.
\newblock Deep neural networks with {L1} and {L2} regularization for high dimensional corporate credit risk prediction.
\newblock {\em Expert Systems with Applications}, 213:118873, March 2023.

\bibitem{lee_graph_2021}
Jong~Wook Lee, Won~Kyung Lee, and So~Young Sohn.
\newblock Graph convolutional network-based credit default prediction utilizing three types of virtual distances among borrowers.
\newblock {\em Expert Systems with Applications}, 168:114411, April 2021.

\bibitem{shen_new_2021}
Feng Shen, Xingchao Zhao, Gang Kou, and Fawaz~E. Alsaadi.
\newblock A new deep learning ensemble credit risk evaluation model with an improved synthetic minority oversampling technique.
\newblock {\em Applied Soft Computing}, 98:106852, January 2021.

\bibitem{xu_efficient_2023}
Biao Xu, Yao Wang, Xiuwu Liao, and Kaidong Wang.
\newblock Efficient fraud detection using deep boosting decision trees.
\newblock {\em Decision Support Systems}, 175:114037, December 2023.

\bibitem{Park_2021}
Min~Sue Park, Hwijae Son, Chongseok Hyun, and Hyung~Ju Hwang.
\newblock Explainability of machine learning models for bankruptcy prediction.
\newblock {\em IEEE Access}, 9:124887--124899, 2021.

\bibitem{Mahajan_2023}
Akshat Mahajan and Kaushal~Kumar Shukla.
\newblock Analyzing false positives in bankruptcy prediction with explainable ai.
\newblock In {\em 2023 International Conference on Artificial Intelligence and Applications (ICAIA) Alliance Technology Conference (ATCON-1)}, pages 1--5, 2023.

\bibitem{bresson2024kagnns}
Roman Bresson, Giannis Nikolentzos, George Panagopoulos, Michail Chatzianastasis, Jun Pang, and Michalis Vazirgiannis.
\newblock Kagnns: Kolmogorov-arnold networks meet graph learning, 2024.

\bibitem{wang2024kolmogorovarnoldinformedneural}
Yizheng Wang, Jia Sun, Jinshuai Bai, Cosmin Anitescu, Mohammad~Sadegh Eshaghi, Xiaoying Zhuang, Timon Rabczuk, and Yinghua Liu.
\newblock Kolmogorov arnold informed neural network: A physics-informed deep learning framework for solving forward and inverse problems based on kolmogorov arnold networks, 2024.

\bibitem{liu2024ikanglobalincrementallearning}
Mengxi Liu, Sizhen Bian, Bo~Zhou, and Paul Lukowicz.
\newblock ikan: Global incremental learning with kan for human activity recognition across heterogeneous datasets, 2024.

\bibitem{yang2024endowinginterpretabilityneuralcognitive}
Shangshang Yang, Linrui Qin, and Xiaoshan Yu.
\newblock Endowing interpretability for neural cognitive diagnosis by efficient kolmogorov-arnold networks, 2024.

\bibitem{handal2024kanopdataefficientoptionpricing}
Rushikesh Handal, Kazuki Matoya, Yunzhuo Wang, and Masanori Hirano.
\newblock Kanop: A data-efficient option pricing model using kolmogorov-arnold networks, 2024.

\bibitem{chen2024sckansformerfinegrainedclassificationbone}
Yifei Chen, Zhu Zhu, Shenghao Zhu, Linwei Qiu, Binfeng Zou, Fan Jia, Yunpeng Zhu, Chenyan Zhang, Zhaojie Fang, Feiwei Qin, Jin Fan, Changmiao Wang, Yu~Gao, and Gang Yu.
\newblock Sckansformer: Fine-grained classification of bone marrow cells via kansformer backbone and hierarchical attention mechanisms, 2024.

\bibitem{bodner2024convolutionalkolmogorovarnoldnetworks}
Alexander~Dylan Bodner, Antonio~Santiago Tepsich, Jack~Natan Spolski, and Santiago Pourteau.
\newblock Convolutional kolmogorov-arnold networks, 2024.

\bibitem{HAN2024}
Di~Han, Wei Guo, Yi~Chen, Bocheng Wang, and Wenting Li.
\newblock Personal credit default prediction fusion framework based on self-attention and cross-network algorithms.
\newblock {\em Engineering Applications of Artificial Intelligence}, 133:107977, 2024.

\bibitem{gong2022hybrid}
Ping Gong, Junguang Gao, and Li~Wang.
\newblock A hybrid evolutionary under-sampling method for handling the class imbalance problem with overlap in credit classification.
\newblock {\em Journal of Systems Science and Systems Engineering}, 31(6):728--752, 2022.

\bibitem{sharma2023credit}
Nityanand Sharma and Vivek Ranjan.
\newblock Credit card fraud detection: A hybrid of pso and k-means clustering unsupervised approach.
\newblock In {\em 2023 13th International Conference on Cloud Computing, Data Science \& Engineering (Confluence)}, pages 445--450. IEEE, 2023.

\bibitem{liu2024kan2}
Ziming Liu, Pingchuan Ma, Yixuan Wang, Wojciech Matusik, and Max Tegmark.
\newblock Kan 2.0: Kolmogorov-arnold networks meet science, 2024.

\end{thebibliography}






\section*{appendix}
\label{sec:appendix}
\subsection*{Hyperparameter Sensitivity Analysis}
\label{subsec:hyperparameters}
The hyperparameters of the model have a significant impact on its performance.
In this subsection, while keeping other hyperparameters constant, we will study the influence of the grid number (grid), optimizer, learning rate, width, and depth of the network of KANs on the model's performance (ROC\_AUC, F1 value), training stability, and training time.
\subsubsection*{Number of Grids (grid)}
For the convenience of research, we simplify the depth and width of the network to the greatest extent, setting width=[10,1], and the remaining hyperparameters are shown in Table~\ref{tab:hyperparameters_grid}.
In this subsection, we let grid take values of 3, 10, 50, and 80 respectively, and observe the values of ROC\_AUC and F1 to analyze the impact of grid on the model's performance. The experimental results are shown in Table \ref{tab:grid_aucf1}.

\begin{table}[ht]
\centering
\caption{Performance metrics for different grid values}
\label{tab:grid_aucf1}
\begin{tabular}{lllll}
\toprule
grid & 3 & 10 & 50 & 80\\
\midrule
ROC\_AUC & 0.8498& 0.8584& 0.8613& 0.8640\\
F1 Value & 0.9673& 0.9675& 0.9675&  0.9675\\
\bottomrule
\end{tabular}
\end{table}
It can be observed that as the grid value gradually increases from 3 to 80, the ROC\_AUC value shows a gradually increasing trend, indicating that the model's ability to distinguish positive and negative samples may improve as the grid value increases. When grid = 3, the ROC\_AUC is 0.8498, and when grid = 80, the ROC\_AUC reaches 0.8640, which is a relatively significant change.
For the F1 value, during the process of the grid value changing from 3 to 80, the F1 value remains basically stable, fluctuating around 0.9675. Only when grid = 3, it is 0.9673, with a very small difference from other values. This indicates that within the range of grid values involved in the experiment, the change in grid has a small impact on the balance between precision and recall of the model. The comprehensive performance of the model in terms of the accuracy and completeness of the prediction results is relatively stable under different grid values. \textbf{And it can be seen that even with the most simplified network configuration such as width=[10,1], by adjusting grid, the model can achieve very good performance.}
\subsubsection*{Optimizer}
In this subsection, let the optimizers be Adam and LBFGS respectively to explore their impact on the training time and model performance. The remaining hyperparameters are shown in Table~\ref{tab:hyperparameters_opt}. The experimental results are shown in Table~\ref{tab:optimizer_comparison}. It can be seen that in terms of training time, Adam has a significant advantage, taking only 9.98s, while LBFGS requires as long as 143.15s. This indicates that in the same hardware environment and other hyperparameter settings, Adam can complete the model training faster. For scenarios with high requirements for training time or the need for rapid iterative experiments, Adam may be a more suitable choice.
In terms of model performance, from the perspective of the ROC\_AUC indicator, LBFGS is slightly better than Adam, with the ROC\_AUC value of LBFGS being 0.8637 and that of Adam being 0.8584. This indicates that LBFGS may be slightly stronger in the ability to distinguish positive and negative samples. However, from the perspective of the F1 value, the two are very close, with the F1 value of Adam being 0.9675 and that of LBFGS being 0.9673. This means that considering both precision and recall, the impact of the two optimizers on the model's performance is not significantly different.
\begin{table}[ht]
\centering
\caption{Performance comparison of different optimizers}
\label{tab:optimizer_comparison}
\begin{tabular}{lccc}
\toprule
Optimizer & Training Time & ROC\_AUC & F1 Value \\
\midrule
Adam & 9.98s& 0.8584& 0.9675\\
LBFGS & 143.15s& 0.8637& 0.9673\\
\bottomrule
\end{tabular}
\end{table}

\subsubsection*{Learning Rate}
\begin{figure}[htbp]
    \centering
    \includegraphics[width=0.8\textwidth]{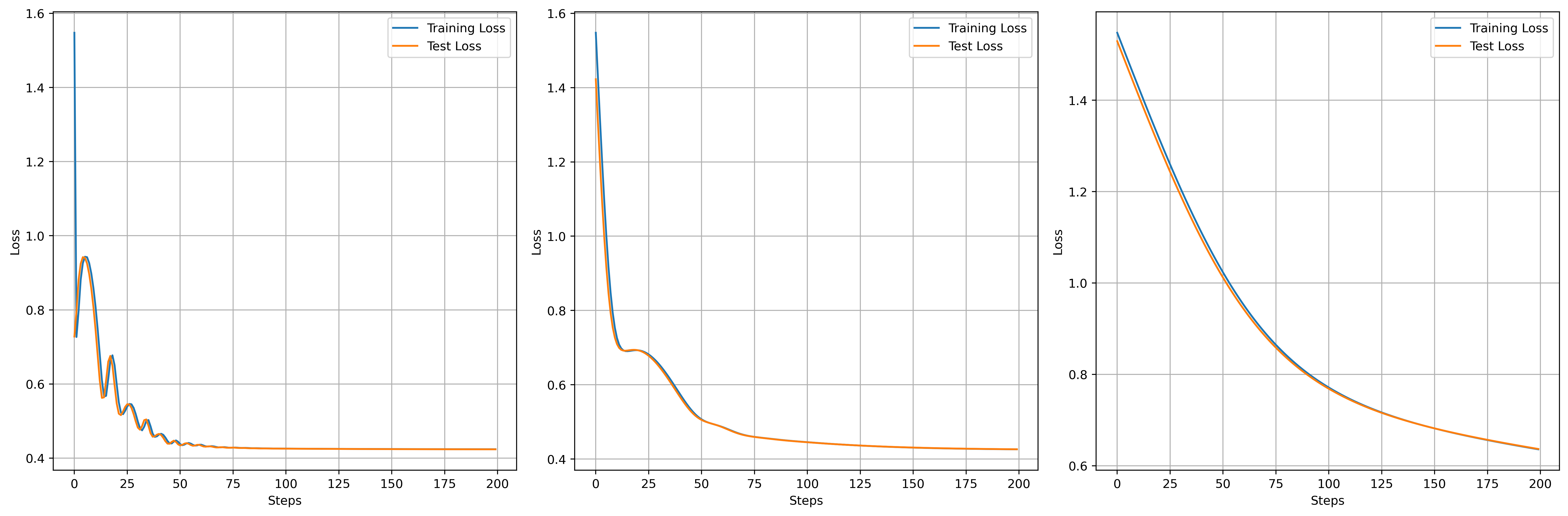}
    \caption{Loss vs. Steps training graph}
    \label{fig:train_lr}
\end{figure}

In this subsection, we aim to deeply investigate the influence of the learning rate on the model's performance and training stability. Specifically, we selected three different values of the learning rate, namely 0.1, 0.01, and 0.001 for experimental exploration. The specific settings of the remaining hyperparameters can be referred to in Table~\ref{tab:hyperparameters_lr}.
By observing Figure~\ref{fig:train_lr}, it can be clearly found that the smaller the learning rate, the more stable the training process. This is because a smaller learning rate makes the step size smaller when the model updates parameters, thereby allowing the model to search towards the optimal solution more smoothly and reducing the possibility of oscillations or deviations from the optimal solution during the training process due to excessive step sizes. However, at the same time, we also note that as the learning rate decreases, the fitting speed of the model becomes increasingly slower. This is because a smaller learning rate means that the amplitude of each parameter update becomes smaller, and the model requires more iterations to achieve a better fitting effect, thereby prolonging the training time.
The results of the impact on the model's performance are shown in Table~\ref{tab:lr_aucf1}. From the data in the table, it can be seen that when the learning rate is 0.1, the ROC\_AUC value is 0.8632 and the F1 value is 0.9674; when the learning rate is 0.01, the ROC\_AUC value is 0.8553 and the F1 value is 0.9672; and when the learning rate drops to 0.001, both the ROC\_AUC value and the F1 value drop significantly to 0.3788. This indicates that within a certain range, an appropriately larger learning rate may contribute to better performance of the model. When the learning rate is too small (such as 0.001), the model's performance will severely decline, possibly because a too small learning rate makes it difficult for the model to effectively learn the features and patterns of the data during the training process and unable to reach a good fitting state within a reasonable time.

\begin{table}[ht]
\centering
\caption{Performance metrics under different learning rates}
\label{tab:lr_aucf1}
\begin{tabular}{llll}
\toprule
Learning Rate (lr) & $0.1$ & $0.01$ & $0.001$ \\
\midrule
ROC\_AUC & 0.8632& 0.8553& 0.3788\\
F1 Value & 0.9674& 0.9672& 0.3788\\
\bottomrule
\end{tabular}
\end{table}

\begin{table}[h]
\centering
\caption{Table for hyperparameter study of KANs (related to grid)}
\label{tab:hyperparameters_grid}
\begin{tabular}{@{}lll@{}}
\toprule
Hyperparameter Name & Value  \\ \midrule
k & 4 \\
batch\_size & -1 \\
optimizer & Adam  \\
learning rate (lr) & 0.1  \\
steps & 100  \\
loss function & \texttt{torch.nn.BCEWithLogitsLoss()} \\ \bottomrule
\end{tabular}
\end{table}

\begin{table}[h]
\centering
\caption{Table for hyperparameter study of KANs (related to optimizer)}
\label{tab:hyperparameters_opt}
\begin{tabular}{@{}lll@{}}
\toprule
Hyperparameter Name & Value  \\ \midrule
width & [10,1] \\
grid & 10 \\
k & 4 \\
batch\_size & -1 \\
learning rate (lr) & 0.1  \\
steps & 100  \\
loss function & \texttt{torch.nn.BCEWithLogitsLoss()} \\ \bottomrule
\end{tabular}
\end{table}

\begin{table}[h]
\centering
\caption{Table for hyperparameter study of KANs (related to learning rate)}
\label{tab:hyperparameters_lr}
\begin{tabular}{@{}lll@{}}
\toprule
Hyperparameter Name & Value  \\ \midrule
width & [10,1] \\
grid & 10 \\
k & 4 \\
batch\_size & -1 \\
optimizer & Adam  \\
steps & 200  \\
loss function & \texttt{torch.nn.BCEWithLogitsLoss()} \\ \bottomrule
\end{tabular}
\end{table}
\end{document}